\documentclass[11pt,draftcls,onecolumn,journal]{IEEEtran}
\usepackage{amsmath,amsfonts}
\usepackage{algorithmic}
\usepackage{algorithm}
\usepackage{array}
\usepackage[caption=false,font=normalsize,labelfont=sf,textfont=sf]{subfig}
\usepackage{textcomp}
\usepackage{stfloats}
\usepackage{url}
\usepackage{verbatim}
\usepackage{graphicx}
\usepackage{cite}
\usepackage{hyperref}
\usepackage{booktabs}
\begin{document}

\makeatletter
\def\@oddfoot{\hbox to\textwidth{\scriptsize\today\hfil\scriptsize Accepted for publication in IEEE Signal Processing Magazine}}
\def\@evenfoot{\hbox to\textwidth{\scriptsize\today\hfil\scriptsize Accepted for publication in IEEE Signal Processing Magazine}}
\def\ps@IEEEtitlepagestyle{%
  \def\@oddfoot{\hbox to\textwidth{\scriptsize\today\hfil\scriptsize Accepted for publication in IEEE Signal Processing Magazine}}%
  \def\@evenfoot{\hbox to\textwidth{\scriptsize\today\hfil\scriptsize Accepted for publication in IEEE Signal Processing Magazine}}%
}
\makeatother
%
\title{An AI Teaching Assistant for Motion Picture Engineering}
%
\author{Deirdre O'Regan, 
Anil C. Kokaram,~\IEEEmembership{Senior~Member,~IEEE}
\thanks{\textcopyright 2026 IEEE.  Personal use of this material is permitted.  Permission from IEEE must be obtained for all other uses, in any current or future media, including reprinting/republishing this material for advertising or promotional purposes, creating new collective works, for resale or redistribution to servers or lists, or reuse of any copyrighted component of this work in other works. Published in \emph{IEEE Signal Processing Magazine}, vol.~43, no.~4, pp.~76--84, July 2026. DOI: \href{https://doi.org/10.1109/MSP.2026.3665423}{10.1109/MSP.2026.3665423}}
\thanks{Deirdre O'Regan and Anil C. Kokaram are with the Department of Electronic and Electrical Engineering, Trinity College Dublin, Ireland. (e-mail: dregan@tcdie; anil.kokaram@tcd.ie).}
\thanks{This work was sponsored by the HEA Human Capital Initiative (HCI) Computational and Data-Centric Engineering Programme (C4-WP4). We also acknowledge the generous support of YouTube to the work of our Sigmedia group.}
}


\maketitle
%
%
%
%
%
\section*{}
\label{sec:abstract}
The rapid rise of LLMs over the last few years has promoted growing experimentation with LLM-driven AI tutors. However, the details of implementation, as well as the benefit in a teaching environment, are still in the early days of exploration. This article addresses these issues in the context of implementation of an AI Teaching Assistant (AI-TA) using Retrieval Augmented Generation (RAG) for Trinity College Dublin's Master's Motion Picture Engineering (MPE) course. We provide details of our implementation (including the prompt to the LLM, and code\footnote{\label{fn:code}\url{https://github.com/sigmedia/ai-teaching-assistant}}), and highlight how we designed and tuned our RAG pipeline to meet course needs. We describe our survey instrument and report on the impact of the AI-TA through a number of quantitative metrics. The scale of our experiment (43 students, 296 sessions, 1,889 queries over 7 weeks) was sufficient to have confidence in our findings. Unlike previous studies, we experimented with allowing the use of the AI-TA in open-book examinations. Statistical analysis across three exams showed no performance differences regardless of AI-TA access ($p>0.05$), demonstrating that thoughtfully designed assessments can maintain academic validity. Student feedback revealed that the AI-TA was beneficial (mean = 4.22/5), while students had mixed feelings about preferring it over human tutoring (mean = 2.78/5).

\section*{Introduction: Embracing The LLM}
\label{sec:motivation}
\IEEEPARstart
{S}{in}ce the rise of transformer-based Large Language Models (LLMs) \cite{vaswani2017attention}, AI chatbots like OpenAI's ChatGPT\footnote{\url{https://chatgpt.com}} have gained widespread popularity among students for academic support. Rather than resisting what has become widespread student practice, educators have an opportunity to integrate AI thoughtfully. By developing course-specific AI Teaching Assistants (AI-TAs), instructors can harness the benefits of LLMs, address generic AI chatbots' limitations, and foster valuable discussions about AI in the classroom.

\subsection*{What Is an AI-TA?}

\begin{figure}[ht]
    \centering
    \includegraphics[width=0.8\textwidth]{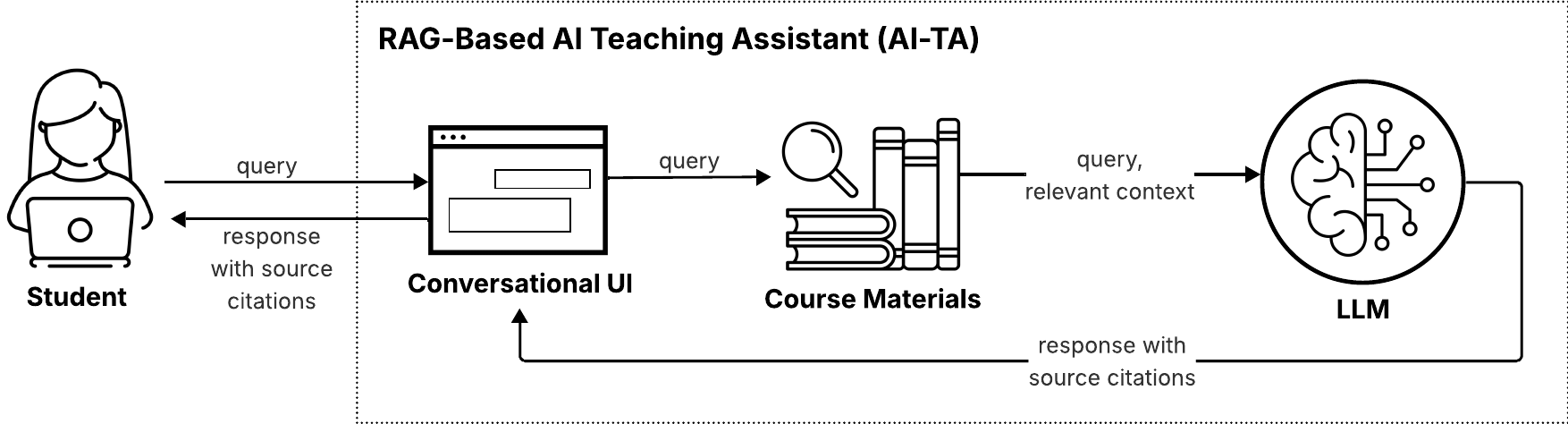}
    \caption{A typical RAG-based AI-TA: Based on the student's query, the AI-TA searches course materials, retrieves relevant context (i.e., excerpts plus source metadata), and forwards the query plus context to an LLM which generates a response grounded in the materials and including source citations.}
    \label{fig:RAG}
\end{figure}

An AI-TA (or AI tutor) is a conversational system customized for a specific course or subject domain. Typically LLM-driven, well-designed AI-TAs provide on-demand support with pedagogical guardrails that promote critical thinking rather than simply dispensing answers. A popular implementation approach employs Retrieval Augmented Generation (RAG)\cite{lewis2020retrieval}, a technique widely adopted for educational AI applications\cite{li2024survey}. RAG addresses a key limitation of LLMs: they can only generate responses based on training patterns, which may be outdated or lack domain-specific knowledge. As Figure~\ref{fig:RAG} shows, the RAG pipeline retrieves relevant excerpts from instructor-curated materials, which the LLM uses to generate responses with source citations. This grounds AI-TA responses in instructor-curated materials rather than generic internet sources, and the citations let students trace AI-generated answers back to authoritative course content, promoting AI literacy alongside knowledge acquisition.

\subsection*{An AI-TA for Motion Picture Engineering (MPE)}

MPE is a Master's course at Trinity College Dublin preparing students for careers in cinematic post-production, visual effects, and video streaming. Building on Digital Signal Processing (DSP) foundations, it covers video manipulation, compression, and transcoding, while exploring neural network-driven advances in the field. Students gain hands-on experience with Foundry's industry-standard Nuke software\footnote{\url{https://foundry.com/products/nuke}} and hear from guest lecturers at Netflix, YouTube, Meta, Sky, VideoLAN, and others. MPE attracts students with a wide technical background, ranging from Deep Learning and DSP practitioners to complete newcomers in both fields. Its assessment structure of take-home assignments and online open-book exams makes it vulnerable to LLM disruption. Instead of resisting this trend, in Spring 2025 we embraced it by deploying a RAG-based AI-TA for MPE (43 students, 296 sessions, 1,889 queries over 7 weeks). 

While several institutions have previously deployed RAG-based AI tutors\cite{neumann2024education, liu2024teaching, liu2024hita, thus2024exploring}, detailed implementation guidance remains limited. This article guides educators through our implementation approach, addressing needs that may resonate—from navigating institutional constraints and pedagogical prompt design to evaluating effectiveness—with code\textsuperscript{\ref{fn:code}} and survey instrument, facilitating replication and adaptation. We investigated: Would students benefit from it? Would they prefer it to a human teaching assistant? What would happen if we allowed its use in an open-book exam?

Student feedback revealed that the AI-TA was beneficial (mean = 4.22/5), and its responses helpful (mean = 4.19/5), while students had mixed feelings about preferring it over human tutoring (mean = 2.78/5). Mixed-methods evaluation uncovered important nuance: students reported positive personal experiences yet expressed concerns about its general use in assessments. Statistical analysis across three open-book exams showed no performance differences regardless of AI-TA access ($p>0.05$), demonstrating that assessments designed specifically for AI-TA access can maintain academic integrity.

Other key implementation insights include: (1) custom AI-TAs provide greatest value when accessing specialized knowledge unavailable in general tools like ChatGPT; (2) Microsoft Azure-based deployments effectively address institutional data governance and compliance requirements; (3) pedagogically informed prompt engineering proved effective in practice; (4) students strategically concentrate usage around assessments rather than maintaining steady engagement; (5) anonymity, while ethically desirable, limits opportunities for personalized support and iterative improvement based on student feedback.
%
%
%
%
%
\section*{RAG `n' Roll: Our Implementation}
\label{sec:research} 

The RAG pipeline has two phases: (1) indexing—source materials are chunked, converted to vector embeddings, and stored in a searchable database; (2) query time—the query is similarly embedded and used to retrieve relevant chunks, which provides context to the LLM. The LLM generates a response grounded in this context, with source citations. This approach offers key educational advantages including domain specificity without LLM retraining, transparency through verifiable citations, and widely purported hallucination\footnote{LLMs ``hallucinate'' when they generate text that sounds plausible but is factually incorrect or unsupported by real data} reduction. We now turn to the specifics of our implementation, highlighting the critical design choices that shaped it.

\subsection*{Platform Selection and Architecture}

While numerous off-the-shelf RAG solutions exist (e.g. OpenAI GPTs, Agenthost, Open-Custom-GPT\footnote{OpenAI GPTs:~\url{https://openai.com/index/introducing-gpts/}, Agenthost:~\url{https://agenthost.ai}, Open-Custom-GPT:~\url{https://github.com/SamurAIGPT/Open-Custom-GPT}}), our particular requirements presented unique challenges: preserving students' anonymity, safeguarding their chat conversations in case of accidental private information disclosure, and complying with Trinity's data protection policies. After careful analysis, we decided to custom-build our RAG-based AI-TA using components from Microsoft Azure and Microsoft Foundry\footnote{Microsoft Azure:~\url{https://azure.microsoft.com/}, Microsoft Foundry:~\url{https://ai.azure.com}}, for the following reasons:

\begin{itemize}
    \item \textbf{Data protection and compliance:} Azure is Trinity's approved enterprise cloud platform. 
    Microsoft's Azure OpenAI service offers Azure-hosted versions of OpenAI's models with configurable content-safety filters, customer-controlled data residency and erasure, and a guarantee that customers' data won't be used for model training.
    \item \textbf{Development and customization:} Microsoft Foundry provides RAG-specific tools and templates that accelerate development, while Azure provides secure infrastructure for hosting a custom-built web client, facilitating iterative refinement based on student feedback.
\end{itemize}

Figure~\ref{fig:architecture} illustrates both our AI-TA's architecture, and its student query workflow. The system comprises five main parts, each involving key implementation decisions:

\begin{figure}[ht]
    \centering
    \includegraphics[width=\textwidth]{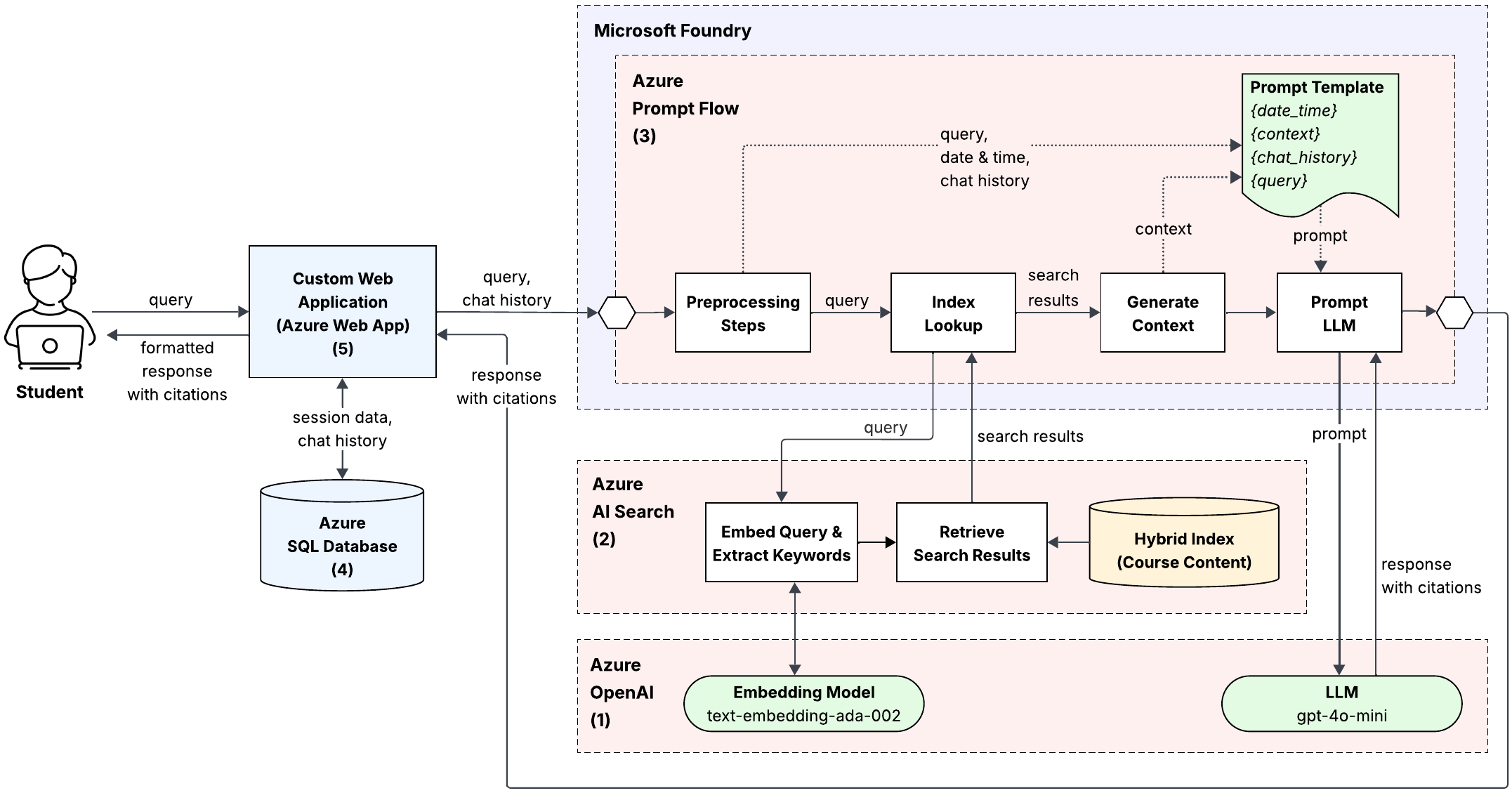}
    \caption{Our AI-TA's implementation architecture and core student workflow. To use this system, create a Project in Microsoft Foundry, and from there: (1) deploy Azure OpenAI models; (2) upload course materials and deploy an Azure AI Search instance to create a Hybrid Index; (3) deploy an Azure Prompt Flow. In Microsoft Azure, deploy: (4) an Azure SQL Database; (5) our custom web application as an Azure Web App. Our web application code, Prompt Flow, Prompt Template, and detailed setup instructions are available on GitHub\textsuperscript{\ref{fn:code}}. You will need to adapt our Prompt Flow and Prompt Template for your specific needs.}
    \label{fig:architecture}
\end{figure}

\begin{enumerate}
    \item \textbf{AI Models:} We chose Azure OpenAI's gpt-4o-mini LLM for its balance of capability and cost-effectiveness, along with Azure OpenAI's text-embedding-ada-002 for vectorization. While an embedding model capable of image understanding may have benefited MPE's visual content, available options required third-party services raising data protection concerns. 
    \item \textbf{Azure AI Search:} Provides ``hybrid + semantic'' search combining vector and keyword-based (i.e., hybrid) retrieval across the indexed course materials, with semantic re-ranking of search results. We adopted Microsoft Foundry's default content chunking settings, prioritizing rapid deployment over extensive optimization. This usefully yielded slide-level citation granularity for lecture slide PDFs. We configured k=10 (retrieve the 10 most relevant results) based on exploratory testing with representative queries, finding this provided good precision-recall balance for MPE's content density.
    \item \textbf{Azure Prompt Flow:} Designed using Microsoft Foundry, this orchestrates the RAG pipeline which manages query processing, context retrieval, Prompt Template population, and LLM inference. Figure~\ref{fig:architecture} shows the Prompt Flow we designed specifically for MPE. Preprocessing steps address two unexpected challenges: (a) content-safety filter false positives (e.g., ``How do I install Nuke'' triggered violent content detection), which we mitigated through phrase replacement;  (b) time-dependent queries (e.g., exam schedules, deadlines) required capturing the current date and time, which we injected into the Prompt Template with instructions for temporal reasoning. Once configured, we deployed the Prompt Flow on an Azure Virtual Machine via Microsoft Foundry, which provided a secure endpoint for application integration.
    \item \textbf{Azure SQL Database:} Stores students' anonymous session data and conversation history, enabling anonymous telemetry to protect student privacy while facilitating cohort-level analysis.
    \item \textbf{Custom Web Application:} Provides anonymous log in, a conventional chat interface with mobile responsiveness, configurable-length conversation history within browser sessions (limited to the 10 most recent exchanges for our MPE deployment), markdown-rendered responses, LaTeX mathematics display, and one-click code copying. This was deployed as an Azure Web App.
\end{enumerate}

\subsection*{Our Prompt and Its Pedagogical Foundations}

Our AI-TA was designed to operationalize elements of dialogic pedagogy\cite{alexander2020dialogic} through explicit prompt instructions emphasizing collaborative knowledge construction. The Prompt Template (available on GitHub\footnote{\url{https://github.com/sigmedia/ai-teaching-assistant/blob/main/tools/azure-prompt-flow-examples/mpe-experiment-2025/chat_with_context.jinja2}}) directed the LLM to ``encourage deeper understanding by posing follow-up questions'' such as ``Are you interested in the mathematical formula or the practical application?'' and to guide students ``step-by-step instead of just stating the answer'' while providing ``worked examples where relevant''. For multi-turn conversations, it instructed the LLM to ``maintain conversation flow by recalling previous interactions'' and ``summarize key points before continuing'' when students built on earlier questions. 

The Prompt Template and web application were designed as a cohesive system to address cognitive load and self-directed learning: the prompt mandated full-path, granular source citations (e.g., slide- or page-level for precise traceability) and markdown/LaTeX formatting, and the application rendered the formatted content faithfully while maintaining conversation history for incremental knowledge construction.

\subsection*{Operational Workflow}

\begin{figure}[ht]
    \centering
    \includegraphics[width=0.95\textwidth]{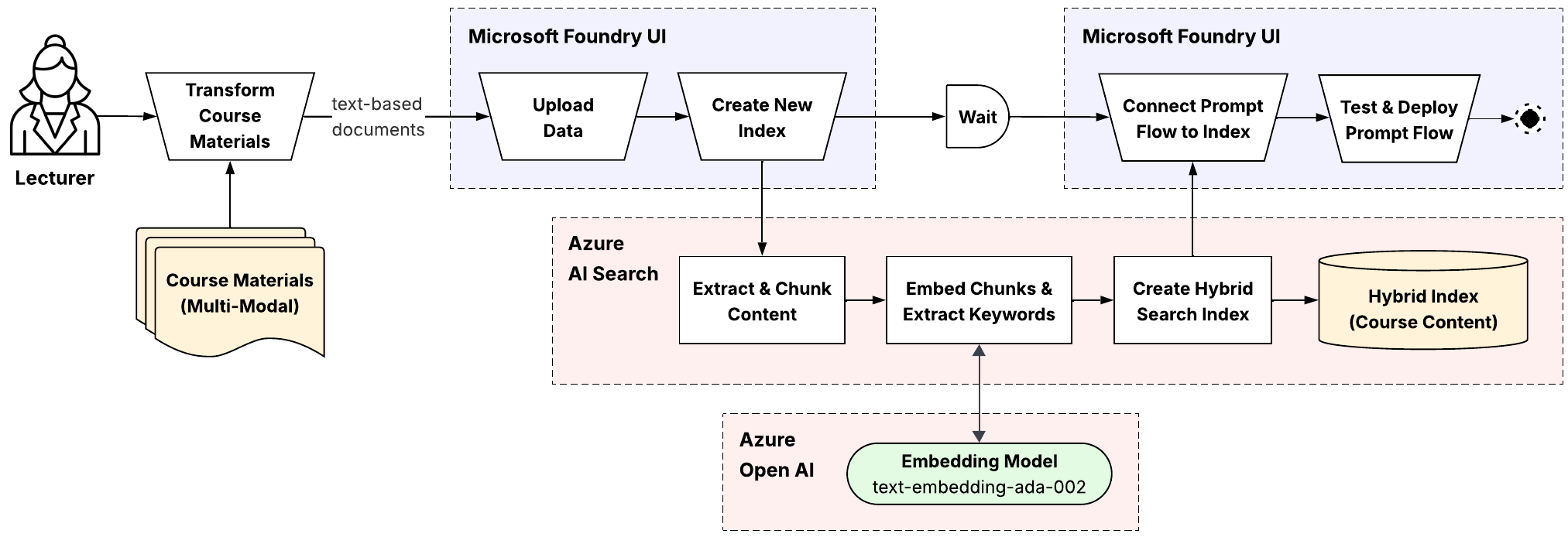}
    \caption{Workflow for ingesting course materials: First, transform all course materials to text-based documents, then using Microsoft Foundry's UI: upload the documents, create a new index, connect the Prompt Flow to the index, and test and deploy the Prompt Flow.}
    \label{fig:operational}
\end{figure}

Figure~\ref{fig:operational} illustrates the lecturer's workflow for ingesting course materials into the AI-TA's Hybrid Index. For MPE, course materials included lecture slides, announcements, Matlab and BlinkScript\footnote{Nuke's proprietary scripting language} scripts, along with audio transcriptions from lectures. Although Spring 2025's MPE lectures were recorded in Microsoft Teams, which has built-in transcription, we used OpenAI's Whisper Turbo\footnote{\url{https://github.com/openai/whisper}} instead, as it appeared to produce better results in informal testing.

%
%
%
%
%

\section*{MPE Deployment: Build It And They Will Come?}
\label{sec:protocol}

For the first 6 weeks of the MPE semester, we prepared the AI-TA, uploading and indexing course materials, conducting exploratory testing and tuning. We launched it to students on week 7 to coincide with Trinity's lecture-free Reading Week. All 43 MPE students received anonymous access via Blackboard\footnote{Trinity's Learning Management System (LMS)} alongside a Participant Information Leaflet (PIL) explaining that AI-TA use was voluntary and anonymous, and that their grades would not be affected by whether they chose to use it or not\footnote{Students provided informed consent by checking a box on the AI-TA's login page acknowledging they had read the PIL, as approved by Trinity's Research Ethics Review Process}. It also cautioned against disclosing private information, and to verify all responses.

\begin{figure}[ht]
    \centering
    \includegraphics[width=\textwidth]{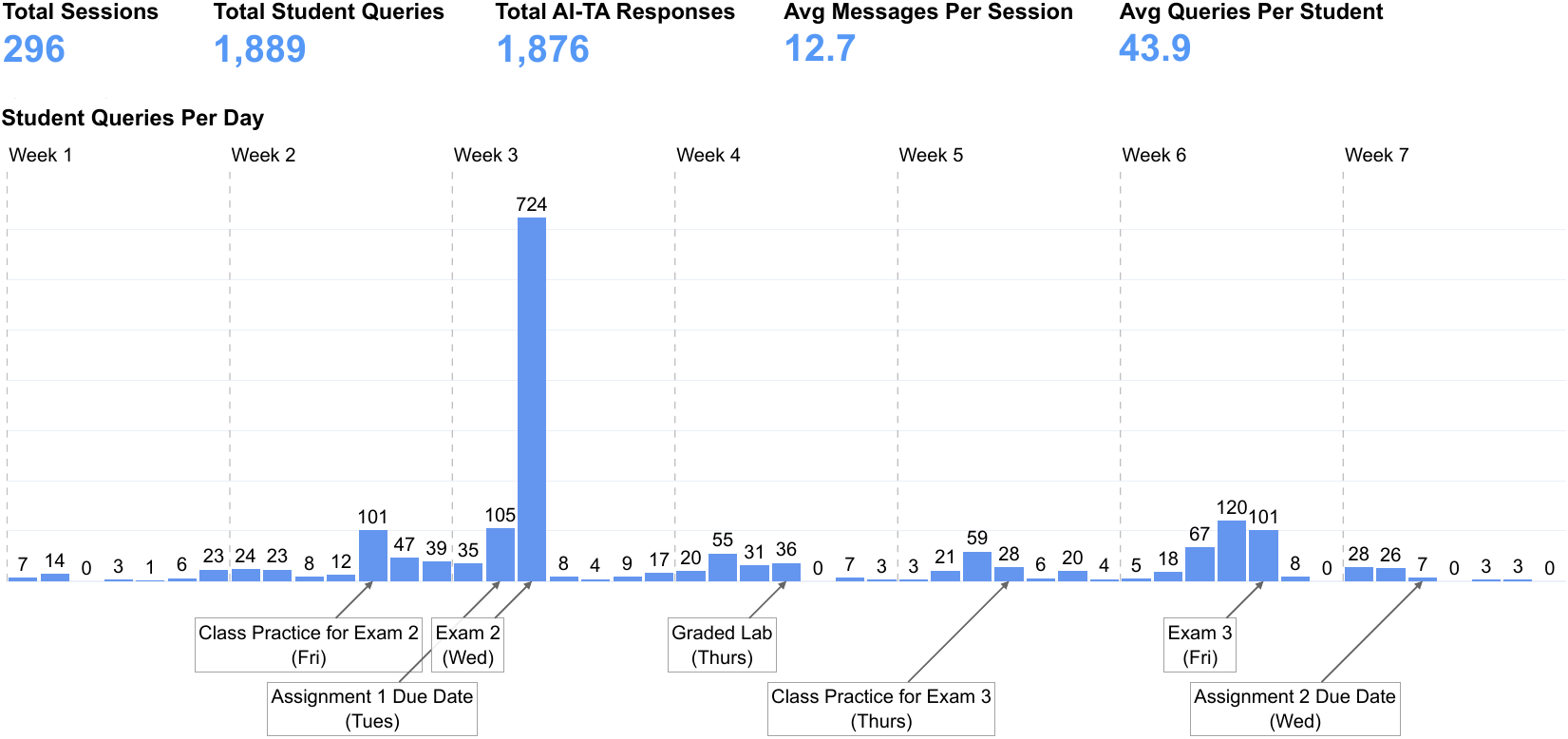}
    \caption{AI-TA engagement: Number of MPE student queries per day over 7 weeks. Annotations show correlations between usage peaks and assessment-related events. Overall engagement statistics are displayed above the chart.}
    \label{fig:usage}
\end{figure}

The AI-TA remained online for 7 weeks until the end of semester and was updated 2–4 times weekly with  lecture materials and transcripts following MPE's delivery schedule. Figure~\ref{fig:usage} shows daily student engagement and overall statistics. 

Initial engagement was low, but usage increased following an in-class reminder (Class Practice for Exam 2), subsequently correlating with assessment deadlines and exams. Assignment 1, which involved implementing video matting techniques in Nuke and reporting on results in an IEEE-style paper, generated queries about report-writing and structuring. For Exam 2, an invigilated open-book quiz delivered via Blackboard, students were permitted to query the AI-TA alongside their notes, but not other AI chatbots (e.g., ChatGPT). The exam period produced a usage spike: 42 active sessions accounting for 564 (78\%) of that day's 724 queries. While session counts don't map directly to individual students, the volume suggests widespread AI-TA use during the exam. In week 6, students sat Exam 3—another Blackboard quiz without AI-TA access but with notes permitted. Preceding days generated queries about likely exam questions, which the system appropriately declined to predict while offering study guidance instead. Assignment 2's deadline in week 7 produced a modest final usage peak. Open-book Exam 1 is not shown in Figure~\ref{fig:usage}, as it was administered prior to the AI-TA's launch.
%
%
%
%
%
\section*{Student Outcomes and Perceptions}
\label{sec:outcomes}

To assess the AI-TA's impact, we examined both student outcomes and perceptions by comparing student performance across three open-book exams administered under different AI-TA availability conditions and analyzing student perceptions through a course exit survey. The study cohort comprised 43 students enrolled in MPE during Spring 2025. Demographics reflected MPE's typical composition: 27 male students (63\%), 16 female (37\%); aged 20–24: 30 (70\%), 25–29: 12 (28\%), 30+: 1 (2\%); prior backgrounds in Engineering: 26 (60\%), Computer Science: 16 (37\%), other fields: 1 (2\%). 36 of the 43 students (84\%) answered the course survey.

\subsection*{Student Performance}

To evaluate whether AI-TA access affected exam performance—a key concern for academic integrity, we compared student scores across three open-book exams (all invigilated quizzes delivered via Blackboard). In all three exams, students could reference their own notes but not other AI chatbots. The three conditions were: Exam 1 (administered before AI-TA launch), Exam 2 (AI-TA permitted during exam), and Exam 3 (AI-TA not permitted during exam). Exam 2 was designed with AI-TA access in mind, emphasizing higher-order thinking skills that could not be directly answered through AI queries alone.

Mean exam scores were: Exam 1 ($\mu=53.5\%$), Exam 2 ($\mu=52.5\%$), and Exam 3 ($\mu=52.3\%$). We performed a paired permutation test with the Student's t-test statistic~\cite{pascutto2011bootstrap}, accounting for individual student differences while controlling for multiple comparisons. All comparisons yielded $p>0.05$ (Exam 1 vs 2: $p=0.94$; Exam 1 vs 3: $p=0.92$; Exam 2 vs 3: $p=0.94$), indicating no significant performance differences across the three exam conditions. These results suggest that with appropriately designed assessments, AI-TA availability does not compromise open-book exam validity.

\subsection*{Survey Results}

The course exit survey comprised 24 mandatory questions, with 6 assessing student perceptions of the AI-TA. Our survey instrument and anonymized data are publicly available on GitHub\footnote{\label{fn:survey}\url{https://github.com/sigmedia/ai-teaching-assistant/tree/main/tools/evaluation/exit-survey/mpe-experiment-2025}}. We compare our results with the MoodleBot study\cite{neumann2024education}, which implemented a RAG-based AI chatbot for a Bachelor's Computer Science course (Database \& Information Systems) at a German university and surveyed 46 participants with a 65\% response rate. We designed three questions specific to our deployment alongside three adapted from their survey. Our students received 1\% of their overall course grade as incentive to complete the survey. Questions Q1–Q5 employed 5-point Likert scales with an additional ``Not Applicable'' option:

\begin{enumerate}
    \item[Q1] I found the AI Teaching Assistant beneficial in supporting my learning in the course.
    \item[Q2] The AI Teaching Assistant's answers to my questions were helpful and informative.
    \item[Q3] The AI Teaching Assistant enhanced my overall experience during the open book exam compared to relying solely on printed notes or memory (i.e. during the second quiz held March 19\textsuperscript{th}, 2025).
    \item[Q4] Given a choice, I would prefer interacting with the AI Teaching Assistant rather than with a real tutor.
    \item[Q5] I would continue to use the AI Teaching Assistant in the future as part of other courses.
    \item[Q6] Tell us more about what you thought of the AI Teaching Assistant. Do you think it should be allowed in open book exams? Did it help you to understand any topics which were otherwise difficult to grasp from the lecturer's explanations? If you didn't use it, why not? Is there anything else you want to mention about it?
\end{enumerate}

Questions Q2, Q4, and Q5 were adapted from the MoodleBot study with key modifications: we added a ``Not Applicable'' (N/A) option, and our Q4 options ranged from ``Never'' to ``Always'' (testing our hypothesis that students' preferences for the AI-TA over a real tutor might be situational). We assigned numerical scores (Strongly Disagree/Never = 1; Strongly Agree/Always = 5) to Q1–Q5 responses to enable quantitative comparison with the MoodleBot study, calculating population mean ($\mu$) and standard deviation ($\sigma$) while excluding N/A responses.

\begin{figure}[ht]
    \centering
    \includegraphics[width=\textwidth]{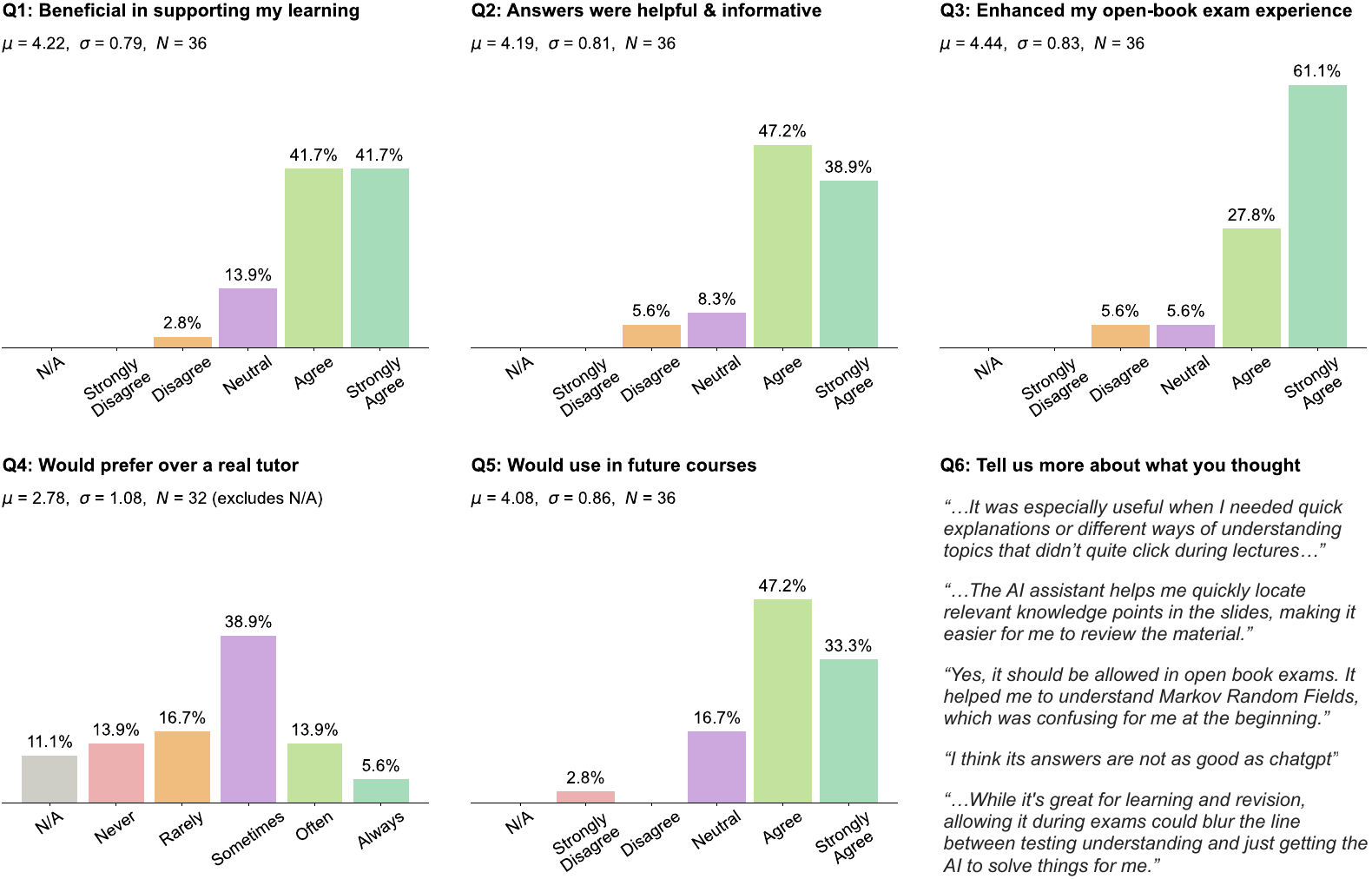}
    \caption{Student perceptions of the AI-TA from the course exit survey (question texts summarized for brevity). We calculated the mean, $\mu$, and standard deviation, $\sigma$, for Q1–Q5. $N$ denotes the number of respondents to each question (excluding ``Not Applicable'' responses). Interesting quotes from open-ended Q6 responses are also shown.}
    \label{fig:survey}
\end{figure}

Results indicated strongly positive perceptions: Q1 ($\mu=4.22, \sigma=0.79, N=36$), Q2 ($\mu=4.19, \sigma=0.81, N=36$), Q3 ($\mu=4.44, \sigma=0.83, N=36$), and Q5 ($\mu=4.08, \sigma=0.86, N=36$). Q4 showed mixed responses ($\mu=2.78, \sigma=1.08, N=32$), suggesting students view the AI-TA as complementary to rather than a replacement for human instruction, appreciating one option or the other depending on the situation. Figure~\ref{fig:survey} visualizes the Likert distributions for questions Q1–Q5. 

Table~\ref{tab:moodlebot_comparison} compares our results with the MoodleBot study for equivalent questions (Q2, Q4, Q5). All differences in means and standard deviations were below 6\%, demonstrating consistency across the two implementations despite differences in course content, institutional, and educational context.

\begin{table}[ht]
\centering
\caption{Comparison of AI-TA with MoodleBot Survey Results}
\begin{tabular}{@{}lccccccc@{}}
\toprule
\textbf{Question (text summarized)} & \multicolumn{2}{c}{\textbf{AI-TA}} & \multicolumn{2}{c}{\textbf{MoodleBot}} & \multicolumn{2}{c}{\textbf{Difference}} \\
\cmidrule(lr){2-3} \cmidrule(lr){4-5} \cmidrule(lr){6-7}
 & $\mu \pm \sigma$ & $N$ & $\mu \pm \sigma$ & $N$ & $\Delta\mu$ (\%) & $\Delta\sigma$ (\%) \\ 
\midrule
Q2: Answers were helpful \& informative & $4.19 \pm 0.81$ & 36 & $4.4 \pm 0.77$ & 30 & -4.8 & +5.2 \\
Q4: Would prefer over a real tutor & $2.78 \pm 1.08$ & 32 & $2.7 \pm 1.14$ & 30 & +3.0 & -5.3 \\
Q5: Would use in future courses & $4.08 \pm 0.86$ & 36 & $4.23 \pm 0.85$ & 30 & -3.5 & +1.2 \\ 
\bottomrule
\end{tabular}
\vspace{5mm}

\raggedright\small\textit{$\Delta\mu$ and $\Delta\sigma$ represent percentage differences in population means and standard deviations respectively.}
\label{tab:moodlebot_comparison}
\end{table}

To extract richer contextual insights from open-ended Q6 responses, we assigned a sentiment rating to each response and tallied approval for AI-TA availability in open-book exams. Student quotes in this article were lightly edited for spelling/grammar (unedited responses are available as supplementary material\textsuperscript{\ref{fn:survey}}). Sentiment distribution was as follows: Positive 30.6\% (11), Positive with Caveats/Reservations 38.9\% (14), Neutral/Balanced/Limited Use 13.9\% (5), and Negative/Skeptical 16.7\% (6). Exam use approval was distributed as follows: Yes 33.3\% (12), No 38.9\% (14), Not Clear 27.8\% (10). 

Students appreciated the AI-TA for clarifying complex concepts not fully grasped in lectures (e.g., Markov Random Fields), and for quickly locating specific information without hunting through slides, validating the usefulness of granular source citations. However, four students noted answer quality concerns ranging from vague responses to incorrect information, while several expressed concerns about dependency, over-reliance, or reduced motivation to learn deeply. Students' cautious stance on exam approval contrasts with Q3 responses where 89\% agreed/strongly agreed that it enhanced their exam experience, reflecting nuanced concerns about broader AI adoption in assessment contexts despite positive individual experiences. Students who expressed reservations worried that AI-TA availability could undermine motivation to master material or provide unfair advantages to less conscientious students, with one candidly noting using it ``mainly so that I'm not disadvantaged by everyone else using it.'' Students proposed mitigation strategies including limiting prompts per student, deducting marks per query, or offering bonus points for non-usage.
%
%
%

%
%
%
\section*{Cross-Study Comparison, Lessons Learned and Recommendations}
\label{sec:lessons}

Our seven-week deployment revealed the educational potential and practical challenges of AI-TAs. We compare our experience with similar systems, distill key lessons learned, limitations, and offer guidance for educators considering similar deployments.

\subsection*{Comparing Implementations Across Contexts}

We compare our AI-TA with five other AI tutors: MoodleBot \cite{neumann2024education}, CS50 Duck \cite{liu2024teaching}, HiTA \cite{liu2024hita}, OwlMentor \cite{thus2024exploring}, and Smeaton \cite{smeaton2025student} across diverse contexts, including undergraduate Computer Science (CS50 Duck, HiTA, MoodleBot), Master's Educational Technology (OwlMentor), undergraduate Business \& Marketing (Smeaton), and Master's MPE (ours). Most deployed during active courses while MoodleBot, developed for a 700+ student course, was evaluated post-course with 46 students who had previously completed it. Cohort sizes varied substantially: CS50 Duck (50,000+ users), HiTA (400), Smeaton (143), MoodleBot (46), our AI-TA (43), OwlMentor (16). 

\textbf{Engagement Patterns and Performance Outcomes:} Usage varied substantially: HiTA averaged 51.9 queries/student over Fall 2023 semester, our AI-TA 43.9 queries/student over 7 weeks, Smeaton 19.5 queries/student over 12 weeks, MoodleBot 3.5 queries/student in post-course evaluation. CS50 Duck reported 50,000+ users with 5–15 prompts per active user daily during Fall 2023, though direct comparison is complicated by its scale and continuous availability across multiple courses. Despite differences in deployment durations and contexts, a consistent pattern emerged: strategic concentration around assessments rather than steady engagement. Our AI-TA received 38.3\% of total queries during one exam, Smeaton documented 26.2\% in 24 hours before the final exam, and OwlMentor showed declining mid-course engagement (150 to 71 events) with renewal before exams (121 events). Our study uniquely examined AI-TA access during assessments, finding no significant performance differences ($p>0.05$) across three exams with varied AI-TA availability. HiTA reported narrowing performance gaps post-deployment, though with potential confounds, while OwlMentor found usage frequency showed no significant correlation with learning outcomes ($p = 0.581$).

\textbf{Student Perceptions:} Positive perceptions emerged across implementations despite varying survey instruments: HiTA (97.9\% beneficial), CS50 Duck (88\% always/frequently helpful), our deployment (83.4\% agreed/strongly agreed the AI-TA was beneficial for learning; 86.1\% agreed/strongly agreed responses were helpful and informative). Our survey's adaptation of MoodleBot questions enabled some direct comparison (see Table ~\ref{tab:moodlebot_comparison}). OwlMentor revealed declining Perceived Usefulness during deployment, with higher self-efficacy students showing lower usage—suggesting AI tools may particularly benefit students lacking initial confidence.

\textbf{AI Model Choices:} Most RAG-based systems converged on OpenAI's GPT family with text-embedding-ada-002 for vectorization. CS50 Duck and our AI-TA hosted models on Azure (CS50: GPT-4; ours: GPT-4o-mini), while others accessed OpenAI's API directly. Smeaton's deployment diverged, adopting Agenthost with incrementally fine-tuned GPT-3.5 instead of RAG (we include this as a notable example of active-course AI-TA deployment using an alternative architectural approach).

\textbf{Costs:} Our 49-day deployment cost approximately EUR~841 (EUR~0.45/query), with EUR~6.23 (EUR~0.003/query) of this attributed to token-based AI usage\footnote{Our costs include some exploratory development with shared infrastructure}. Therefore, fixed-tier Azure infrastructure (virtual machine, search engine, web application plan, database, etc) dominated our expenses. Post-deployment analysis suggests that we over-provisioned the virtual machine, with at least 40\% cost reduction possible through right-sizing infrastructure in future deployments, with further economies possible through system tuning and extension for multi-course deployment. Only CS50 Duck and MoodleBot reported costs for comparison: CS50 Duck reports \$0.05/query (cost components not detailed); MoodleBot reports \$0.01–\$2.33/message accounting for API token costs only.

\subsection*{Other Lessons Learned}

\textbf{Competing With Commercial Chatbots:} Despite significant prompt engineering and retrieval tuning effort, informal testing often revealed ChatGPT's responses to be as good as our AI-TA's. ChatGPT is improving rapidly and is now extremely difficult to beat on public knowledge, even for specialized topics like MPE. However, RAG-based AI-TAs excel in two scenarios: (1) private institutional knowledge (e.g., ``when is my 3rd quiz scheduled?''), and (2) recent specialized information that commercial chatbots have not yet incorporated. For instance, our AI-TA knew updated FFmpeg\footnote{\url{http://ffmpeg.org}} command-line syntax from course materials while ChatGPT did not, which was critical since students needed this for their assignments. 

\textbf{Pros and Cons of Anonymity:} Smeaton, OwlMentor, and our AI-TA employed usage anonymity. Smeaton found students asked questions they wouldn't ask in lectures, with similar benefits noted in our survey where one student wrote: ``Sometimes I have loads of dumb questions, that's when it becomes more helpful.'' This judgment-free space was encouraging given students' reticence to speak up in class this year. However, anonymity also prevented collaborative discussions on AI-TA features and improvements, and made targeted individual support impossible. Future work will explore whether aggregate query analysis can identify challenging concepts for in-situ teaching adaptation or iterative course refinement. 

\subsection*{Limitations}

\textbf{Technical:} Our text-only RAG implementation prevented image understanding, which would have been valuable for MPE's visual content. We relied on exploratory testing rather than formal accuracy evaluation, expecting that evaluation by the lecturer for a few key queries was sufficient. Post-deployment analysis revealed that low-context queries from mid-conversation (e.g., ``what is the main reason'') yielded poor context retrieval results. The operational workflow for updating course materials required approximately 1 hour per update, representing significant administrative effort given 2–4 updates weekly for MPE.

\textbf{Study Design, Biases and Comparability:} Our quasi-experimental design lacked random assignment and control groups, precluding strong causal inferences that would be possible with a randomized controlled trial. Our small scale (43 students) limits statistical power versus large implementations (CS50 Duck: 50,000+ users). Our survey used positively-worded statements (e.g., ``enhanced my open-book exam experience'') likely inflating agreement rates—a limitation shared with MoodleBot and OwlMentor. Our 84\% response rate may reflect incentive bias (1\% grade for participation), while anonymous access prevented correlating usage with outcomes. Our Master's-level cohort (63\% male; 70\% aged 20–24; 97\% Engineering/Computer Science) limits generalizability. Deployment context differences further complicate cross-study comparisons: our in-course deployment versus MoodleBot's post-course volunteer evaluation involve distinct participant motivations and selection effects.

\subsection*{Recommendations for Implementers}

\textbf{Platform Selection, Cost Optimization, and Data Privacy:} Clarify your institution's data governance requirements early, as these determine available options. With flexible data policies, off-the-shelf platforms like Agenthost (used by Smeaton~\cite{smeaton2025student}) may provide richer features and reduced operational burden. For stricter data protection or residency (e.g., GDPR compliance), Azure-hosted solutions may be preferable, and costs can be optimized by right-sizing fixed-tier services. Regardless of platform choice, consider user-facing privacy measures. Analysis of 100 random AI-TA queries revealed students avoided entering personally identifiable information; our system displayed ``Do not disclose any information which could identify an individual,'' suggesting clear privacy notices effectively encourage data caution.

\textbf{Anonymity Trade-offs:} Consider anonymity's pedagogical and practical implications. While a judgment-free environment may encourage questions students wouldn't ask publicly, anonymity complicates individual follow-up and targeted support while preventing correlation of usage with outcomes.

\textbf{Evaluation and Assessment Design:} Employ experimental designs with control groups when feasible, and use mixed-methods evaluation combining quantitative and qualitative data. Report participant demographics and deployment contexts to enable meaningful cross-study comparisons. When permitting AI-TA access during assessments, design questions requiring critical evaluation, higher-order thinking, or responses difficult to generate via AI prompting.
%
%
%
%
%
\section*{Conclusion}

This article demonstrates that educators can successfully deploy RAG-based AI-TAs by addressing institutional, pedagogical, and technical considerations, with our implementation code and survey instrument\textsuperscript{\ref{fn:code}} enabling replication and adaptation. Statistical analysis across three open-book exams revealed no significant performance differences regardless of AI-TA access ($p>0.05$), demonstrating that thoughtfully designed assessments can maintain academic integrity. Students rated the AI-TA highly beneficial (mean = 4.22/5), though expressed nuanced concerns about broader assessment use despite positive personal experiences. We plan to extend our AI-TA to additional Engineering courses with improvements including operational workflow automation, formal accuracy evaluation, assessment usage protocols, prompt flow improvement for low-context queries, and aggregate query analysis to inform adaptive teaching. The rapid evolution of LLM capabilities suggests exciting opportunities ahead for educators willing to engage thoughtfully with these technologies.

%
%
%
%
%
\section*{Acknowledgments}
Many thanks to Trinity's John Squires for support with Microsoft Foundry, to Mark Linanne, Vibhoothi, Amir Bijedic, and Harun \v{S}iljak for technical and ethics support, to Evelyn Fox and Colm Lawlor for data protection recommendations, and to the Sigmedia group for testing the AI-TA.
%
%
%
%
%

\end{document}